# Using Neural Network to Propose Solutions to Threats in Attack Patterns

*[1]Adetunji Adebiyi, [2]Chris Imafidon
[1,2] School of Architecture, Computing and Engineering, University of East London
[1]adetunjib@hotmail.com, [2]C.O.Imafidon@uel.ac.uk

*Abstract.* In the last decade, a lot of effort has been put into securing software application during development in the software industry. Software security is a research field in this area which looks at how security can be weaved into software at each phase of software development lifecycle (SDLC). The use of attack patterns is one of the approaches that have been proposed for integrating security during the design phase of SDLC. While this approach help developers in identify security flaws in their software designs, the need to apply the proper security capability that will mitigate the threat identified is very important. To assist in this area, the uses of security patterns have been proposed to help developers to identify solutions to recurring security problems. However due to different types of security patterns and their taxonomy, software developers are faced with the challenge of finding and selecting appropriate security patterns that addresses the security risks in their design. In this paper, we propose a tool based on Neural Network for proposing solutions in form of security patterns to threats in attack patterns matching attacking patterns. From the result of performance of the neural network, we found out that the neural network was able to match attack patterns to security patterns that can mitigate the threat in the attack pattern. With this information developers are better informed in making decision on the solution for securing their application.

* Corresponding Author:
Adetunji Adebiyi
10, Blenheim Road
Leighton Buzzard, UK,
Email: [1]*adetunjib@hotmail.com*   Tel: +447848377803

## 1. Introduction

The development of secured software application in today's hostile environment is increasingly challenging. Many software applications are now running critical mission systems where security of these systems cannot be ignored. However, tight deadlines for software delivery and the increase in the skills of attackers in uncovering vulnerabilities in software applications makes it more difficult in building software with adequate level of security assurance [16]. While reliance on network security has been effective in protecting software in the past, it is no longer adequate as attackers now aim at attacking vulnerable software applications directly [12].

With this in mind, a lot of effort is now put into building security into software application during SDLC in the software industry. In line with this view, software security research field emerged in the last decade and views security as an emergent property of software application under development and not a feature to be bolted on to the application after development. As a result, it has been proposed that security needs to be weaved into software application all through SDLC [13]. Previous research shows that most security problems in software applications are caused by defects or design flaws in the software [4] [7] [14]. Therefore, by integrating security into software during the early stages of software development, security flaws will be addressed early when it is less costly and easier to fix than during the later stages [5] [16] [20].

To achieve this, a neural network based tool is proposed in this paper that will match possible attack patterns in the design of a software product to security patterns that is able to mitigate the risk in the attack patterns. Neural network based applications has been used successfully in the area of network security as intrusion detection systems, misuse detection systems and firewalls [23] [25]. Also in the field of application security, neural network has been proposed to be used in technologies such as virus detection system, authentication system and cryptography [24] [26]. It would be noticed however, that these neural network based applications can only provide a form security after software deployment. Its usage in this paper is during the design stage of software development before software is coded.

Various techniques have been proposed for integrating security into software products during SDLC. Attack patterns and security patterns are examples of these techniques and the proposed neural network tool complements these techniques when integrating security into software design. Attack patterns describe the methods through which an attacker may use to hack into a software product [5]. With the aid of attack patterns, developers are able to view the objectives of the attacker or think like the attacker in order to anticipate the threat to their software product and subsequently integrate appropriate security capability [1]. Security pattern captures security solutions to recurring security problems and this can be used by developers in making knowledgeable decisions on the security of their software application [19].

Over the years, a lot of security patterns has been developed and has been classified using different criteria. Therefore, developers intending to use security patterns as part of their strategy in integrating security into their software products are faced with a new problem of finding the security pattern that adequately address the security problem in their software. This is especially true for developers lacking the knowledge of security requirements of their software under development and its potential threats. As most developers lack this knowledge, many software development projects involve security experts to assist in integrating security. While the involvement of security experts in software development projects is seen as a good security practice [14] [17], it also creates conflict between the developers and security experts because of the existing gap between them [27]. In these projects, developers are often burdened with fixing security defects identified by the security experts which are not their primary concern as they have been trained to think of functions and features of their product and its on-schedule delivery [9]. Under this situation developers are forced to make a choice between functionality or releasing their products to the market and security [6].

However, with the proposed neural network based tool in this paper, developers especially lacking the necessary security skills can match possible attack patterns that have been identified in their software design to security patterns that can provide appropriate security solution to threats in the design. Based on this, developers can take the necessary steps in integrating the security patterns that will help in making their software application secure. Apart from this, the gap between security experts and developers reduces significantly as developers are guided in making informed decisions about the security of their software applications.

## 2. Attack Pattern

An attack pattern is a description of how an attacker can carry out an attack that occurs in a particular situation. In one of the earliest occurrence of attack patterns in literature, Moore 2001 defined it as "as a generic representation of a deliberate, malicious attack that commonly occurs in specific contexts" [15]. Moore stated that each attack pattern contains an overall goal of the attacker, a list of preconditions, sequence of steps for staging the attack and a list of post of conditions. From the pioneer work of Christopher Alexander on design patterns, the idea on attack patterns was conceived by security experts [1]. Design patterns capture solutions to reoccurring software design problems in object oriented program [22]. Following the pattern construct of design patterns, the representation of attacker's perspective during an attack is now presented in the form of attack patterns. By describing how an attack can be executed in various contexts, attack patterns enable developers identify vulnerabilities in their software and various ways for exploiting the vulnerabilities [5]. Also, just like problem-solution paradigm of design patterns, attack patterns also provide recommendation for mitigating the attacks [1].

In a similar way to security pattern the intention of attack patterns is to enable software developers who lack the experience in software security to learn how to determine vulnerabilities and possible attacks in their software products [22]. As a result, attack patterns have been created by security experts and researchers and they serves as a means of transferring the essential knowledge needed for building secure software to software developers. With this knowledge, developers will learn how their software can be exploited and how they can implement various security capabilities to mitigate the attacks [1].

Furthermore, attack patterns help to reduce the knowledge gap between software developers and attackers [1]. While software developers run against time and limited budget to develop quality software products, attackers have all the time they need to exploit targeted software application. Also, to make their software secured, developers must envisage different ways their software can be attacked and implement adequate security capabilities to prevent these attacks. However, attackers only need to find out only one point of weakness in the target software product to carry out their exploit and today there are a lot of sophisticated tools at their disposal to carry out such exploits [3][11]. As the skills of the attackers continue to increase, it has become more challenging for the software developers to keep abreast the knowledge attackers have acquired in exploiting software [1]. Therefore, with the use of attack patterns, the knowledge gap reduces significantly as the developers learn to think like the attacker to anticipate how the security of their software can be compromised.

Today, many attack patterns have been created by different authors in the software industry. Example of these include 48 attack patterns created by Hoglund and McGraw, 2004 and 51 regularly expressed attack patterns by Gegick and Williams, 2006. However, due to conflicts in definitions and structures of attack patterns, an effort to provide some coherence has been put together by the U.S Department of Homeland Security (DHS) sponsored repository, CAPEC (Common Attack Pattern Enumeration Classification) where attack patterns instances are collected and made available to the public [1].

There are some related techniques to attack patterns which are also useful in helping software developers in evaluating their software product from an attackers' point of view in order to find potential threats and plan mitigation. Attack trees is one of these techniques and it is used to characterize system security by modeling the decision making process of attackers. In this technique, attack against a system is represented in a tree structure in which the root of the tree represents the goal of an attacker. The nodes in the tree represent the different types of actions the attacker can take to accomplish his goal on the software system or outside the software system which may be in the form of bribe or threat [5] [18]. Attack net is another technique which includes "places" analogous to the nodes in an attack tree. This is used to indicate the state of an attack. Events required to move from one place to the other are captured in the transitions and arcs connecting places and transitions indicate the path an attacker takes. Similarly, vulnerability tree is another technique which is a hierarchy tree constructed based on how one vulnerability relates to another and the steps an attacker has to take to reach the top of the tree [18].

## 3. Research Overview

The neural network tool proposed in this paper suggests possible security solutions in form of security patterns to security flaws that has been identified in the software design of an application based on attack pattern that captures the security flaw in question. Figure 1 is an overview on the process of mapping attack patterns to security patterns by the neural network tool. From the software design, possible attack patterns are identified during security analysis of the software design. Information capturing the identified attack pattern is used to model the attack pattern which is then evaluated by the neural network tool. Based on the evaluation of the attack pattern, the neural network tool then suggests possible security patterns that can mitigate the security threats in the attack pattern. With this information, the developer can proceed to apply the appropriate security capability to their software design

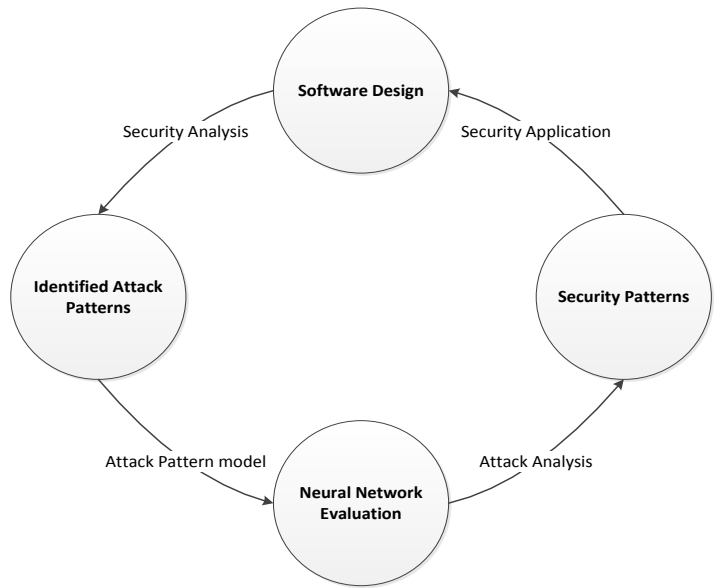

**Figure 1.** Mapping attack pattern to security patterns

This approach builds on the proposed selection criterion by Wiesauer and Sametinger 2009 which matches attack patterns to security design patterns. By using their proposed taxonomy the authors argue that developers can match identified attack patterns in their software designs to corresponding security design patterns that would provide the appropriate mitigation. Based on this approach, the proposed neural network tool is trained to map security patterns to attack patterns using information abstracted from attack patterns. William and Gegick 2006 proposed 51 regular expression-based attack patterns which help in indicating the sequential events that occur during an attack. These attack patterns are based on the software components involved in an attack and are used for identifying vulnerabilities in software design. Information on the components forming the attack patterns was abstracted and used as attributes of the attack pattern. This attack pattern attributes is then used to train the proposed neural network tool. The attributes consist of the following:

The Attack ID: This is the unique ID that identifies the attack
Resource Attacked: This is the resource that is attacked in the attack pattern.

Attack Vector: This is the method through which the attacker uses to attack the resource
Attack Type: This state whether the attack is an attack against confidentiality, integrity or availability

Furthermore, in order to group the attack patterns into similar threat category, Microsoft threat classification scheme (STRIDE) was used to categorize the attack patterns into six groups. Figure 2 shows that out of the 51 regularly expressed attack patterns, 1 of them was classified under spoofing identity attack category, 2 was classified under tamper with data attacks, none was classified under repudiation attacks, 6 was classified under the information disclosure attacks, 21 was classified under the denial of service attacks and 27 was classified under the elevation of privilege attacks. No attack was classified under repudiation attacks because none of the regularly expressed attack patterns demonstrated this type of attack. However, it was assumed that this attack was covered under the elevation of privilege attack because the attacker must have escalated his privileges before been able to cover his tracks in a multi-stage attack scenario.

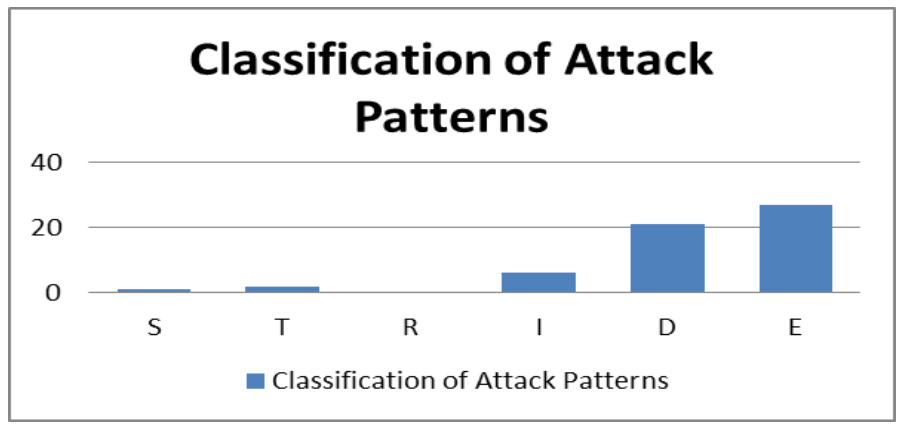

**Figure 2.** Number of attack patterns classified according to STRIDE

Data was also collected from the security design patterns defined by Steel, et.al (2005), Blakley, et.al (2004) and Kinezle and Elder (2003). A total of 23 security design patterns were defined by Steel, et.al (2005). These were classified into four logical tiers consisting of the web tier, the business tier, web services and identity tier. A total of 13 security design patterns were defined by Blakley, et.al (2004) and these were classified into two groups. This consisted of the available security design patterns and the protected security design patterns. The security design patterns defined by Kinezle and Elder (2003) were also classified into two categories. These include the structural patterns and procedural patterns. The structural patterns consist of 13 main security design patterns and 3 mini-patterns. The mini-patterns are less formal and shorter discussion that were included as a supplement to the main security design patterns. The procedural patterns consist of 13 security design patterns defined for the purpose of improving the development process of mission critical software applications. As these security patterns are not applied in the software application, they were not used as part of the data for training the neural network.

Based on the above, the following is the limitation of this research approach. Firstly, the solution proposed by the neural network tool is limited to attacks patterns possible due to security flaws caused by software design defects. Secondly the security solution proposed by the neural network tool is based on security patterns which address web application security issues. Thirdly, the neural network tool has been trained based on information from attack patterns proposed by William and Gegick 2006 only. Some of these limitations will be addressed in the future as discussed in section five.

### 3.1 The Neural Network Architecture

The standard three layer neural network architecture consisting of the input layer, the hidden layer and the outer layer was adopted in training the neural network. A feed-forward back-propagation neural network is used to analyse the attack patterns and generate possible solutions from the security design patterns that can be help in mitigating the threat identified in the attack patterns. Resilient back propagation algorithm was applied to optimize the performance of the neural network. A tan-sigmoid transfer function was also applied to the various connection weights in the hidden nodes and output nodes.

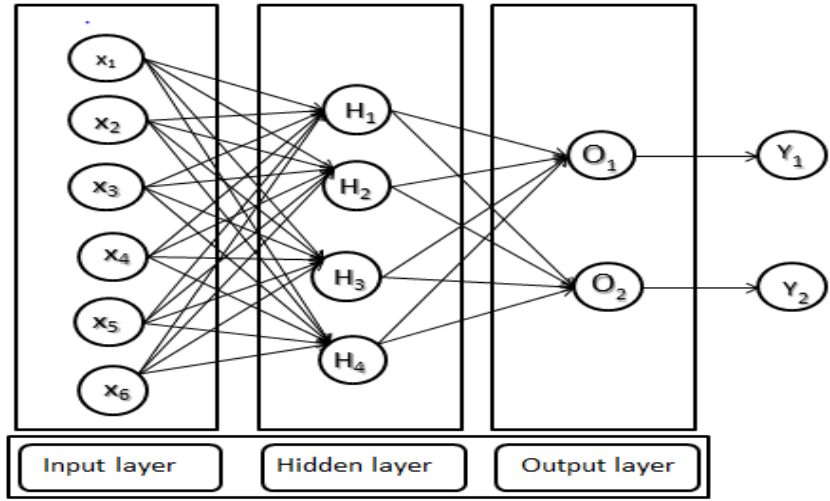

**Figure 3.** Neural Network Architecture

### 3.2 Data Encoding

Using the attributes of the attack patterns discussed above the collected data was analysed and encoded. Table 1 shows the attributes of the regularly expressed attack patterns, the observed data for each attribute and the value assigned to each attribute when it is observed.

**Table 1.** Attributes of Regularly Expressed Attack Patterns

| s\no | Attribute | Observable | Value |
|---|---|---|---|
| 1 | Attack ID | Attack Pattern | Attack ID |
| 2 | Resource Attacked | Attack Component | Attack Component ID |
| 3 | Attack Vector | Attack Component | Attack Component ID |
| 4 | Attack Type | Availability | 1 |
|   |   | Integrity | 2 |
|   |   | Confidentiality | 3 |

Following the analysis of the data collected on the regularly expressed attack patterns the data needed for training the neural network was encoded. A total of 226 training data samples were abstracted from the regularly expressed attack patterns using the attributes in Table 1. To encode the data, the corresponding value for the information abstracted by each attribute in the Table was used in encoding the data. For instance, regularly expressed attack pattern 1 is represented as:

$$(User^+)(Server^+)(Log^+)(HardDrive^+)$$

By using the attributes in Table 1 for the regularly expressed attack pattern the information on Table 2 was collected. In this attack pattern, the attacker (user) uses the log as an attack vector for crashing the hard drive (which is the target of the denial of service attack) by causing the server to log large amount of data in the log file in order to fill the hard drive when the server is flooded with a lot of requests to be processed.

Table **2**: Sample of Pre-processed training data from attack pattern

| Attack ID | Resource Attacked | Attack Vector | Attack Type |
|---|---|---|---|
| 1 | Hard Drive | Log | Availability |

In order to encode the information abstracted in the Table 2, the attack component ID for Hard Drive and Log was used for their encoding. The corresponding value for Availability in Table 1 was also used for its encoding. Table 6, shows the training data for the example above after it has been encoded.

Table 3: Sample of training data after encoding

| Attack ID | Resource Attacked | Attack Vector | Attack Type |
|---|---|---|---|
| 1 | 42 | 58 | 1 |

The next stage involves converting the encoded data into ASCII comma delimited format which can be used to train the neural network as shown below

1, 42, 58, 1

The data is then loaded into the neural network for training as shown in the following Table.

Table 4: Sample of input data into neural network

| Input 1 | Input 2 | Input 3 | Input 4 |
|---|---|---|---|
| 1 | 42 | 58 | 1 |

For the expected output, security design patterns by Blakley, et.al (2004), Steel, et.al (2005), and Kinezle and Elder (2003) were grouped into six groups with respect to STRIDE based on an approached used by Halkidis, S.T. et al. (2006). In this approach, security patterns by Blakley, et.al (2004) was analysed qualitatively using Microsoft threat classification (STRIDE) to find out the security design pattern that provides protection on each of the threat category. Therefore each group provides possible solutions to the threats identified under each threat category of STRIDE. A unique ID is assigned each group so that the neural network can match them to the corresponding attack patterns. Based on this encoding, the neural network is expected to identify the possible solution for the attack pattern by giving the following output:

1, 0, 0, 0, 0, 0

### 3.3 Neural Network Training

To train the neural network the training data set is divided into two sets. The first set of data is the training data (201 Samples) that was presented to the neural network during training. The second set (26 Samples) is the data that was used to test the performance of the neural network after it had been trained. The training performance is measured by Mean Squared Error (MSE) and the training stops when the generalization stops improving. Mat lab Neural Network tool box was used to perform the training. The training parameters also include the learning rate which is set to 0.01 with a goal of 0; maximum fail set to 6 and a minimum gradient of 0.000001. To find the best performance of the neural network with the lowest MSE, different numbers of neurons were applied to the network. The training was executed in five simulations for each number of neurons the applied to the network to obtain the average results of its performance because the neural network is initiated with random weights during its training and this gives different results. Therefore, the average results on the training time, MSE and number of epoch were used in analysis of the performance of the neural network.

## 4. Result and Discussion

During the training of the neural network, it was observed that the highest MSE (0.002286) was obtained when 110 neurons were applied in the network and the lowest (0.001055) was obtained when 90 neurons were applied. It was also observed that the highest number of epoch used is 1324.8 when 80 neurons were applied and this decreased as the number of neurons applied increased. The lowest number of epoch used was 906.8 when 110 neurons were applied. With respect to the time spent in training the neural network, it was observed that the shortest time spent in training (25 seconds) was when 90 neurons were applied in the network and the longest time spent in training the network (61 seconds) was when 80 neurons were applied to the network. It was noticed that the time spent in training the network decreased as the number of neurons applied to the network increased from 80 to 110 and increased slightly when 120 neurons were applied. The performance of neural network was also tested using the test data. Table 5 shows the actual and expected output of the neural network

Table 5: Actual and expected output of Neural Network

| s\n | Test Data Sample | Actual Output | Expected Output |
|---|---|---|---|
| 1 | Sample 1 | 6.0000 | 6 |
| 2 | Sample 2 | 5.9999 | 6 |
| 3 | Sample 3 | 5.0000 | 5 |
| 4 | Sample 4 | 4.9998 | 5 |
| 5 | Sample 5 | 5.0000 | 5 |
| 6 | Sample 6 | 5.0000 | 5 |
| 7 | Sample 7 | 5.9999 | 6 |
| 8 | Sample 8 | 6.0000 | 6 |
| 9 | Sample 9 | 5.0000 | 5 |
| 10 | Sample 10 | 2.6441 | 6 |
| 11 | Sample 11 | 5.0000 | 5 |
| 12 | Sample 12 | 5.0000 | 5 |
| 13 | Sample 13 | 6.0000 | 6 |
| 14 | Sample 14 | 4.9183 | 5 |
| 15 | Sample 15 | 6.0000 | 6 |
| 16 | Sample 16 | 5.0000 | 5 |
| 17 | Sample 17 | 6.0000 | 2 |
| 18 | Sample 18 | 1.7707 | 2 |
| 19 | Sample 19 | 5.6890 | 6 |
| 20 | Sample 20 | 4.0000 | 4 |

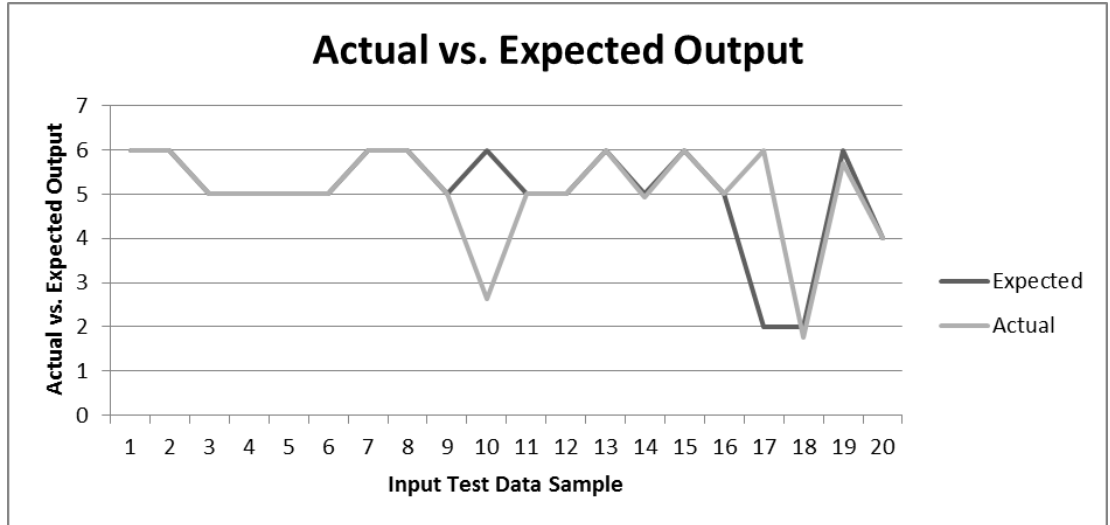

**Figure 3:** Actual vs. Expected output

By comparing the expected and actual output, it would be seen that the network was able to match most of the attack patterns to the expected group that will provide mitigation to vulnerabilities in the attack pattern. However, there were two instances in which the network failed to match the attack patterns to the expected group. This was when the network was used to evaluate test data sample 10 and 17. For test data sample 10, the network produce an output of 2.6441 when the expected output is 6 and for test data 17, the network produced an output of 6 when the expected output is 2. By looking at the data used in training the network for matching the attack patterns to their corresponding security pattern, it was seen that for these attack patterns, the attacker had multiple ways in which the attack could be carried out. This explains why the network failed to match the attack patterns. With a larger data sample for training the neural network, a better performance can be achieved. Figure 3 is a graph showing the difference between the actual and expected output

## 5. Future Work

While the neural network tool is able to match attack patterns to groups of security patterns that can mitigate the threat in the attack patterns, it is our intention to look further into improving the output result of the neural network tool by using other suggested classification of security patterns when defining the expected output of the neural network. Furthermore, for each group of security patterns that is currently used in the expected output, we hope to design test cases that can be used to validate whether the suggested security patterns can provide mitigation to the attack patterns. A comparative analysis between our approach and other approaches for integrating security into software during development would be carried out. Also, further work is required to train the network to match attack patterns to other security design patterns because the neural network has been trained to match attack patterns to three security design patterns. In addition, the neural network needs to be thoroughly tested before it can gain acceptance as a tool for matching attack patterns to security design patterns.

## 6. Conclusion

This paper has demonstrated how neural network can be used has a tool to propose solutions to threats in attack patterns by suggesting possible security patterns that can provide mitigation. While there may be challenges in developing secured software application, the hostile environment in which critical mission applications are expected to run today makes security non-negotiable. The necessary skills for making software applications secured have been limited to the field of security experts in the past. However, the use of attack patterns and security patterns are part of current approaches which software developers lacking knowledge on security can adopt in order to develop secure software applications. Therefore, with the aid of the proposed neural network tool in this paper, developers identifying particular attack patterns in their software application can easily be pointed to security patterns that will be resolve the security problem in their application. And by integrating the necessary security capabilities based on the information from the output of the neural network tool, developers will be able to develop secured software application.